\documentclass[12pt, a4paper]{article}

\usepackage{amsmath,amssymb,amsfonts}
\usepackage{graphicx}
\usepackage[colorlinks,hypertexnames=false]{hyperref}
\usepackage{cite}

\begin{document}
\title{Constraints on the neutrino extension of the Standard Model and baryon asymmetry of the Universe}
\author{Volodymyr Gorkavenko${}^1$,
Oleksandr Khasai${}^2$,\\
Oleg Ruchayskiy${}^3$,
Mariia Tsarenkova${}^{1}$\vspace{0.5em}
\\
${}^1$ \it \small Faculty of Physics, Taras Shevchenko National University of Kyiv,\\
\it \small 64, Volodymyrs'ka str., Kyiv 01601, Ukraine\\
${}^2$ \it \small Bogolyubov Institute for Theoretical Physics, National Academy of Sciences of Ukraine,\\
\it \small 14-b, Metrolohichna str., Kyiv 03143, Ukraine\\
${}^3$ \it \small Niels Bohr Institute, University of Copenhagen,
\it \small Blegdamsvej 17, 2010, Copenhagen, Denmark
}
\date{}

\maketitle
\setcounter{equation}{0}
\setcounter{page}{1}%

\begin{abstract}
    Heavy neutral leptons (HNLs) can cause new effective interactions of Standard Model particles, particularly charged lepton flavour violation (cLFV) processes. The non-observation of cLFV processes, therefore, puts constraints on the parameters of the HNLs. We find the relations between the cLFV effective operators in the realistic case when active neutrino masses are non-zero and masses of the HNLs are non-degenerate. This allows us to strengthen the existing cLFV constraints. We also link the baryon asymmetry of the Universe to the same cLFV effective operators, which imposes new restrictions on their values.
\end{abstract}

\section{Introduction}\label{sec:intro}

The Standard Model (SM) effectively describes the electroweak and strong interactions of elementary particles and has been validated up to $\sim$15 TeV in collider experiments \cite{Cottingham:2007zz}. However, phenomena such as neutrino masses \cite{Bilenky:1987ty,Strumia:2006db,deSalas:2017kay}, dark matter \cite{Peebles:2013hla,Lukovic:2014vma,Bertone:2016nfn}, and baryon asymmetry of the Universe \cite{Steigman:1976ev,Riotto:1999yt,Canetti:2012zc} remain unexplained.  
These phenomena demonstrate the incompleteness of the SM and indicate the existence of new hidden particles.

In this paper, we consider neutrino extension of the SM with right-handed (RH) or sterile neutrinos. 
Adding one sterile neutrino to the SM causes only one active neutrino to become massive. This contradicts data on active neutrino oscillations. Adding two sterile neutrinos to the SM causes two active neutrinos to become massive and one neutrino is massless, which is compatible with neutrino oscillation data. 
The most interesting case occurs in the Neutrino Minimal Standard Model ($\nu$MSM) \cite{Asaka:2005,Asaka_2:2005}, when three RH neutrinos are added
\begin{equation}\label{g1}
    \delta \mathcal{L} = i \bar N_I \partial_\mu\gamma^\mu N_I - F_{\alpha I} \bar L_\alpha N_I \Phi - \frac{M_I}2 \bar N^c_I N_I + h.c.
\end{equation}
Here, $N_I$ ($I = 1,2,3$) is RH neutrino,  $\alpha $ corresponds flavours of leptons ($e, \mu, \tau$),  $\Phi$ and $L_\alpha$  are the Higgs and lepton doublets, respectively, $F_{\alpha I}$ are elements of the Yukawa matrix, $M_I$ are Majorana masses of RH neutrinos. 

The appropriate choices of  18 new parameters of the $\nu$MSM can solve the above mentioned three problems of SM. In this case, the lightest RH neutrino is a dark matter particle with a mass
of $\sim\!10$ keV. Two other RH neutrinos are heavy particles with almost equal masses. They
 provide generation of baryon asymmetry in the Universe through the mechanism of leptogenesis \cite{Akhmedov:1998qx,Buchmuller:2004nz,Pilaftsis:2005rv,Davidson:2008bu,Pilaftsis:2009pk,Shaposhnikov:2009zzb,Bodeker:2020ghk}.  Results of \cite{Asaka:2005} were later revised for the case of two \cite{Klaric:2020phc,Klaric:2021cpi} and three \cite{Drewes:2021nqr} RH neutrinos. In particular, it was shown
 that the condition of almost equal masses of the RH neutrinos for baryogenesis is not necessary \cite{Drewes:2012ma}. 

Since we will be interested in the experimental search
for HNL particles, it makes sense to consider a model
with only two heavy RH neutrinos, as the lightest
RH neutrino (dark matter candidate) has a sufficiently smaller coupling to SM particles. 

In this paper, we consider a relation between experimentally observable parameters of the neutrino extension of the SM. The upper bounds on the Lagrangian parameters follow from collider experiments.  We want to consider a lower bound on these parameters that follow from 
 the baryon asymmetry of the Universe in the $\nu$MSM and compare lower and upper bounds among themselves.

\section{Theoretical relations between observable parameters $S_{\alpha\beta}$ and $R_{\alpha\beta}$}\label{sec:Relations}

The parameters of the Lagrangian \eqref{g1} that can be measured in collider experiments are derived as follows \cite{Coy:2018bxr}
\begin{align}
    S_{\alpha\beta} &=\sum_I F_{\alpha I}F^\dagger_{I \beta} M_I^{-2}, \label{Sab}\\
    R_{\alpha\beta} & =\sum_I F_{\alpha I}F^\dagger_{I \beta} M_I^{-2} \ln\frac{M_I}{M_W}.\label{Rab}
\end{align}

The parameters \eqref{Sab} and \eqref{Rab} are limited by upper bounds, which come from non-observation of processes that violate lepton number (e.g., $Z \to e\mu$, $\mu \to e\gamma$, etc), or from measurement errors of observed SM processes (e.g., $Z \to  \ell^+ \ell^-$, etc).
The upper bounds from experiments for $\hat S_{\alpha\beta}=M_W^2 S_{\alpha\beta}$, $\hat R_{\alpha\beta}=M_W^2 R_{\alpha\beta}$, where $M_W$ is the mass of the $W$-boson, are shown in Tabl.\ref{tab:my_label1}

\begin{table}[]
    \centering
 \begin{tabular}{ |c|c|c| } 
 
\hline
          & Present experiments & Future experiments \\
\hline
 quantity &  upper limit  & upper limit \\
\hline
 $\hat S_{ee}+\hat S_{\mu \mu}$ & $0.53\cdot 10^{-3}$ & -\\ 
\hline
 $\hat S_{\tau \tau}$ & $0.64\cdot 10^{-3}$  & - \\
\hline
$|\hat S_{e \mu}|$ & $6.8\cdot 10^{-6}$ &  $2.6\cdot 10^{-6}$ \\
\hline
 $|\hat S_{e \tau}|$& $4.5\cdot 10^{-3}$ & $1.8\cdot 10^{-3}$ \\
\hline
$|\hat S_{\mu \tau}|$&  $5.2\cdot 10^{-3}$  & $1.4\cdot 10^{-3}$ \\
\hline
 $|\hat R_{e \mu}|$ &$2.4\cdot 10^{-7}$ & $1.7\cdot 10^{-8}$ \\
\hline
 $|\hat R_{e \tau}|$  & 0.022 & $3.0\cdot 10^{-3}$\\
\hline
 $|\hat R_{\mu \tau}|$ & 0.019 & $4.2\cdot 10^{-3}$\\
\hline
\end{tabular}
\vspace{1em}    \caption{ Upper bounds on the seesaw parameters $\hat S_{\alpha\beta}$ and $\hat R_{\alpha\beta}$ from present and expected in the foreseeable future experiments, see details in \cite{Coy:2018bxr}.}
    \label{tab:my_label1}
\end{table}

For further analysis it's necessary to express observables $ S_{\alpha\beta}$, $ R_{\alpha\beta}$ via $\nu$MSM parameters, namely:
\begin{itemize}
\item active neutrino oscillations parameters (mixing angles $\theta_{ij}$, masses of active neutrinos $m_\alpha$);
\item  masses of sterile neutrinos $M_I$; 
\item  parameter $U_{tot}^2$, which can be expressed as
$ U^2_{tot} =\sum_{\alpha,I} |\Theta_{\alpha I}|^2= \frac{v^2}{M^2}{\rm tr}(FF^\dagger) = \frac{\sum_i m_i}{M}\cosh2\mathfrak{Im}\omega$, 
where $M$  stands for the mass of the right-handed neutrinos if $M_1 \approx M_2 \approx M$, $\Delta M/M \ll 1 $,  and $\Theta_{\alpha I}$ represents the mixing angle between left-handed neutrinos ($\nu_\alpha$) and right-handed neutrinos ($N_I$), $\omega$ is a complex angle of the Casas-Ibarra parameterization \cite{Casas:2001sr}.
\end{itemize}

In the following we will focus on a region that is of interest for current experiments, where heavy neutral leptons (HNL) are being explored above the seesaw threshold:
\begin{equation}\label{BigOmegaAssum}
    \cosh2\mathfrak{Im}\omega \approx \sinh2\mathfrak{Im}\omega \approx \frac{\exp{2\mathfrak{Im}\omega}}{2}{\gg 1}.
\end{equation}
This assumption should hold for different masses of heavy neutrinos $M_I$ too.

In the extended Standard Model Lagrangian \eqref{g1}, the elements of the Yukawa matrix $F_{\alpha I} $ can be conveniently described using parameters related to active neutrinos through the Casas-Ibarra parameterization \cite{Casas:2001sr}.
Since this parameterization is quite complicated for further calculations and analytical operations, we introduce a new 
$3 \times 3$ complex matrix  $X=\frac{i}{v}U_\nu\sqrt{m_\nu^{\rm diag}}$ to simplify the calculations. Using this new matrix, we can derive relatively simple expressions for observables $S_{\alpha \beta}$ and  $R_{\alpha \beta}$.

For normal ordering of active neutrino masses, we obtain
\begin{equation}\label{SabNorm}
  S_{\alpha \beta}=\frac{e^{2 \mathfrak{Im}\omega}}{4}\frac{ M_1+M_2}{ M_1 M_2}(X_{\alpha2}-i X_{\alpha3}) \left(X_{\beta2}^*+i X_{\beta3}^*\right),
 \end{equation}
\begin{align}\label{RabNorm}
      R_{\alpha \beta}=\frac{e^{2 \mathfrak{Im}\omega}}{4}&\frac{ (M_1\ln\frac{M_2}{M_W}+M_2\ln\frac{M_1}{M_W})}{ M_1 M_2}
      (X_{\alpha2}-i X_{\alpha3}) \left(X_{\beta2}^*+i X_{\beta3}^*\right).
 \end{align}
 As one can effortlessly see, diagonal elements $S_{\alpha\alpha}$ are real and positive.  
Corresponding relations for the case of the inverted active neutrino mass hierarchy can be obtained by replacing second indexes in elements of the $X$ matrix $3 \to 2$, $2 \to 1$.
 
One can derive a relationship between experimentally measurable quantities, specifically the elements of the matrices  $S_{\alpha \beta}$ and $R_{\alpha \beta}$, using Exps. \eqref{SabNorm}, \eqref{RabNorm}:
\begin{equation}\label{SandR}
    S_{\alpha \beta}\Bigl(\!M_1\ln\frac{M_2}{M_W}\!+\!M_2\ln\frac{M_1}{M_W}\!\Bigr)\!=\!R_{\alpha \beta}(M_1\!+\!M_2).
\end{equation}
We can obtain powerful restrictions for observable parameters $S_{\alpha \beta}$ and $R_{\alpha \beta}$ by utilising explicit forms of expressions \eqref{SabNorm}, \eqref{RabNorm}:
\begin{equation}\label{SRrelations}
    |S_{\alpha \beta}|^2= S_{\alpha \alpha} S_{\beta \beta} , \quad
    |R_{\alpha \beta}|^2 =  R_{\alpha \alpha} R_{\beta \beta}.
\end{equation}
These new expressions hold true even when the active neutrinos are massive, and the masses of the two right-handed (RH) heavy neutrinos differ. The only necessary conditions are that the assumption in \eqref{BigOmegaAssum} remains true and that the active neutrino masses are much smaller than the RH neutrino masses.

These constraints \eqref{SRrelations} are significant because previous similar constraints, such as the saturated Schwarz inequality, were derived only in specific cases of the massless active neutrinos and the heavy sterile neutrinos with degenerated masses \cite{Coy:2018bxr,Blennow:2023mqx}. 

If we assume that the conditions in \eqref{SRrelations} hold true with sufficient accuracy above the seesaw line under assumption \eqref{BigOmegaAssum}, the upper limits on the seesaw parameters $\hat S_{\alpha\beta}$ and $\hat R_{\alpha\beta}$ can be further refined, namely
\begin{equation}\label{improved_v}
    |\hat S_{e \tau}| \leq 0.58\cdot 10^{-3}, \qquad |\hat S_{\mu \tau}| \leq 0.58\cdot 10^{-3}.
\end{equation}

\section{Baryon Asymmetry and observable parameters $ S_{\alpha\beta}$ and $ R_{\alpha\beta}$}\label{sec: Baryon}

The question of how baryon asymmetry in the early Universe can be generated in the $\nu MSM$ was considered in detail in \cite{Asaka:2005}. 
Based on \cite{Asaka:2005}, we expressed the baryon asymmetry in terms of the experimentally measurable elements of the matrices $S_{\alpha\beta}$ and $R_{\alpha\beta}$.

In the case of the normal active neutrino mass hierarchy, we have
\begin{align}\label{DeltaB_FormNO}
  &\frac{n_B}{s}  \leq \frac{7 \cdot 10^{-4}\pi^{\frac32}\sin^3\phi}{384\cdot12^{\frac13}\Gamma(\frac56)} \left(\frac{M}{\Delta M_{21}}\right)^{\frac{2}{3}} 
  \frac{M_0^{\frac73} M^{\frac{11}{3}}\sqrt{m_3^2+4m_2^2+8m_3 m_2}}{T_W v^2 M_W^4 } \sum_{\alpha,\beta\neq\alpha}\hat S_{\alpha \alpha}|\hat S_{\alpha \beta}|,
\end{align}
that is valid if 
$M_1\approx M_2 \approx M$. Here $m_i$ are the masses of active neutrinos. 
 For the inverted active neutrino mass hierarchy, we need to substitute masses $m_3 \to m_2$, $m_2 \to m_1$ in the previous equation \eqref{DeltaB_FormNO}. 

We would also like to note that to obtain expression \eqref{DeltaB_FormNO} we need to use precise expressions for $S_{\alpha \beta}$, $R_{\alpha \beta}$ without assumption \eqref{BigOmegaAssum}. It can be shown that the baryon asymmetry of the Universe is proportional to $\mathfrak{Im}[S_{\alpha \beta}^* R_{\alpha \beta}]$, which under condition \eqref{BigOmegaAssum} gives zero.
 
Taking experimental constraints on the elements of the $\hat R$ and $\hat S$ matrices from Table and its improved values \eqref{improved_v}, we get
in the case of the normal hierarchy
\begin{equation}\label{DB1}
   \frac{n_B}{s} \leq 4.6\left(\frac{M}{\Delta M_{21}}\right)^{\frac{2}{3}}(M/1 {\rm GeV})^\frac{11}{3},
\end{equation}
and in the case of inverted hierarchy:
\begin{equation}\label{DB2}
   \frac{n_B}{s} \leq 10.2\left(\frac{M}{\Delta M_{21}}\right)^{\frac{2}{3}}(M/1 {\rm GeV})^\frac{11}{3}.
\end{equation}

Nevertheless, these expressions have limited practical value. The right side is much larger than 1 (${M}/{\Delta M_{21}}\gg 1$ and $M \gtrsim 1$ GeV), while the left side is much smaller than 1 (${n_B}/{s}\sim 10^{-10}$). This discrepancy suggests that the actual values of the elements of the $\hat R$ and $\hat S$ matrices are likely many orders of magnitude smaller than the experimental limits presented in Tabl.\ref{tab:my_label1}.

\section{Conclusions}\label{concl:results}

In this work, we examined an extension of the Standard Model (SM) by adding two heavy right-handed (RH) neutrinos with masses much higher than the electroweak scale. Detecting these heavy neutrinos directly is extremely difficult. But these new particles can theoretically generate charged lepton flavour violating (cLFV) processes. The fact that these cLFV processes haven't been observed helps us place limits on the parameters of heavy neutrinos (HNLs).

We have analytically obtained relations between the observable parameters of active neutrinos and parameters of the neutrino extension of SM (with two heavy RH neutrinos of different masses), as shown in equations \eqref{SandR} and \eqref{SRrelations}. For these relationships to hold, the active neutrinos must be extremely light compared to the RH neutrinos. Also, assumption \eqref{BigOmegaAssum}, important for current experimental searches for heavy neutral leptons, must hold.

We concluded that Schwarz inequalities $|S_{\alpha \beta}|^2 \leq S_{\alpha \alpha} S_{\beta \beta} $ and $|R_{\alpha \beta}|^2 \leq R_{\alpha \alpha} R_{\beta \beta} $ only become equalities (saturate) when $e^{\mathfrak{Im}\omega}\gg 1$. This result is independent of the mass difference between sterile neutrinos, which is crucial because the case of nearly identical right-handed neutrino masses is only a theoretical conjecture. This conjecture is used to explain the Universe's baryon asymmetry in the Minimal Neutrino Model ($\nu$MSM) \cite{Asaka:2005, Asaka_2:2005}. However, as shown in \cite{Drewes:2012ma}, the explanation can also be derived in models where right-handed neutrinos have different masses. It should be noted that previous similar relationships were only derived for particular cases, such as equal heavy sterile neutrino masses or active neutrinos with zero mass \cite{Coy:2018bxr, Blennow:2023mqx}. 

In addition, we strengthened the upper limits on the observable parameters $\hat S_{\alpha\beta}$ and $\hat R_{\alpha\beta}$ by an order of magnitude, as presented in \eqref{improved_v}, assuming that the restrictions in \eqref{SRrelations} hold under the condition in \eqref{BigOmegaAssum}.

We derived an expression for the baryon asymmetry of the Universe using measurable parameters (the elements of the $\hat S$ and $\hat R$ matrices) \eqref{DeltaB_FormNO}. We discovered that under assumption \eqref{BigOmegaAssum}, the baryon asymmetry is equal to zero. 
Only taking $\cosh2\mathfrak{Im}\omega \neq \sinh2\mathfrak{Im}\omega$ into account, a non-zero baryon asymmetry arises.

We demonstrated that the lower limits (baryon asymmetry) and the upper limits (particle accelerator experiments) on the observable $\nu$MSM parameters differ by many orders of magnitude. This indicates that if the baryon asymmetry is really caused by heavy right neutrinos, the actual values of the observed elements of the $\hat S$ and $\hat R$ matrices are much lower than the experimental limits given in Tabl.\ref{tab:my_label1}.

\section{Acknowledgments}

The authors would like to thank the Organizers of the "New Trends in High-Energy and \mbox{Low-x} Physics" Conference in Sfântu Gheorghe (Romania) for their generous hospitality during this extremely interesting and inspiring meeting.
The work of V.G. and O.Kh. was supported by the National Research Foundation of Ukraine under project No. 2023.03/0149.

\end{document}